\documentclass[letterpaper]{article}
\usepackage{iccc}

\usepackage{times}
\usepackage{helvet}
\usepackage{courier}
\usepackage{graphicx}
\usepackage{url}
\usepackage[hidelinks]{hyperref}

\makeatletter
\def\@lbibitem[#1]#2{\item\if@filesw
{\def\protect##1{\string ##1\space}\immediate
\write\@auxout{\string\bibcite{#2}{#1}}}\fi\ignorespaces}
\makeatother

\makeatletter
\renewcommand{\thebibliography}[1]{%
  \section*{References\@mkboth{REFERENCES}{REFERENCES}}%
  \list{}{%
    \setlength{\labelwidth}{0pt}%
    \setlength{\leftmargin}{1.5em}%
    \setlength{\itemindent}{-\leftmargin}%
    \itemsep .01in}%
  \def\newblock{\hskip .11em plus .33em minus .07em}%
  \sloppy\clubpenalty4000\widowpenalty4000%
  \sfcode`\.=1000\relax}
\makeatother

\newcommand\blfootnote[1]{%
  \begingroup
  \renewcommand\thefootnote{}\footnote{#1}%
  \addtocounter{footnote}{-1}%
  \endgroup
}

\title{The Affordance is the Message: Creative Media as Complex Systems}
\author{Ane Espeseth$^{1}$ {\normalfont and} Elias Najarro$^{2}$\\
$^{1}$Center for Computing in Science Education, University of Oslo, Oslo, Norway\\
$^{2}$Play, Culture and AI, IT University of Copenhagen, Denmark\\
\texttt{anekes@uio.no}, \texttt{enaj@itu.dk}\\
}
\begin{document}
\maketitle
\blfootnote{17th International Conference on Computational Creativity (ICCC 2026), Coimbra, Portugal.}

\begin{abstract}
\begin{quote}
The affordances of a creative medium strongly condition the creative artefacts the medium will produce. 
In this work, we present a formalisation of computational creativity (CC) media using the conceptual toolbox of complex systems (CS). We introduce the notions of emergence, collective intelligence and self-organisation, non-linear dynamics, criticality, multi-scale hierarchy, phase transitions, diversity of attractors, path dependence, and open-endedness, and connect them to the existing CC literature. Together these nine properties form a vocabulary with which creative media can be described and compared at the system level, while medium affordances are the design-level mechanisms that determine each medium's complex system properties. 
The formalisation emphasises the influence of each medium's affordances in determining what the medium can produce in creative processes. 
To demonstrate the proposed theoretical approach, we characterise a diverse set of media (\texttt{r/place}, \texttt{Minecraft}, \texttt{cellular automata}, \texttt{Twitter}, \texttt{Boids}, \ldots) using this vocabulary. 
The proposed formalisation under the CS concepts serves a dual purpose: it establishes a link between the affordances and realised practices of the medium, and it offers a shared lens for characterising existing creative media.

\end{quote}
\end{abstract}

\section{Introduction}

A computational medium used to elicit and express creativity can be seen as complex systems: McCormack \shortcite{McCormack2012} argues in his work on \textit{creative ecosystems} that the environments in which creative activity unfolds emerge from interactions among agents, substrates and shared rules. Murray~\shortcite{murray2011inventing} treats the digital medium as an authored bundle of expressive \textit{affordances} ~\cite{gibson1979ecological}: combinations of its surfaces and elements, perceivable and meaningful relative to a particular agent. In this paper, we use the term ``creative medium'' to describe a computational substrate with a set of affordances that hosts creative activity. These affordances combine to produce the system-level properties that characterise the medium as a complex system.

\begin{figure}[t]
  \centering
  \includegraphics[width=\columnwidth]{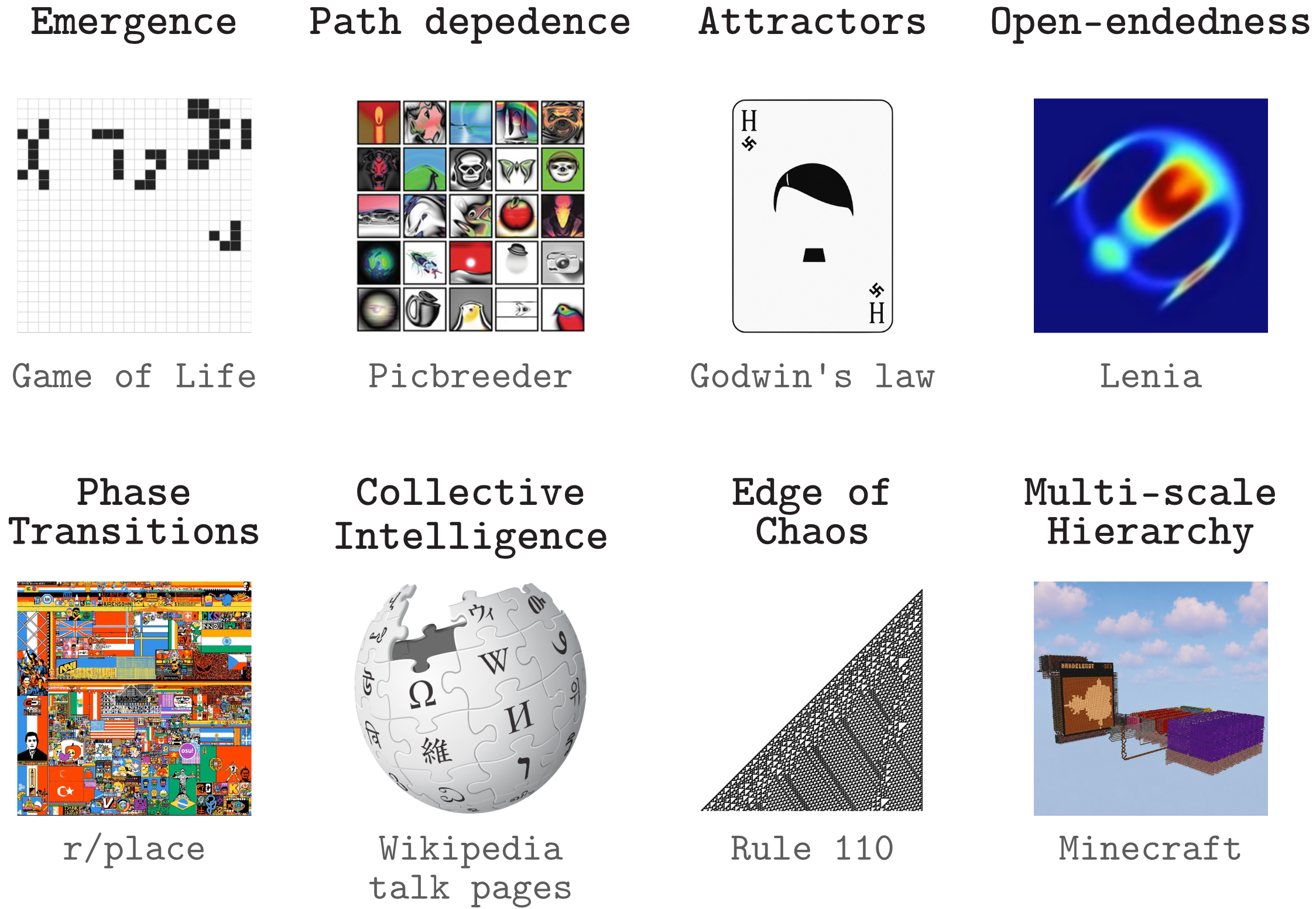}
  \vspace{-3ex}
  \caption{Computational creative media can be studied as complex systems. Complexity science offers a vocabulary of such properties --- a principled basis on which to characterise media and investigate how their affordances condition the creative artefacts they produce. This figure illustrates eight of these properties, each paired with a creative medium that exemplifies it.}
  \label{fig:iccc}
\end{figure}

An illustrative example is the \texttt{r/place} experiment, started on Reddit on April Fools' Day 2017~\cite{rappaz2018latent}. Users acted within a small affordance set: a 1000$\times$1000 pixel canvas, a 16-colour palette, a single-pixel brush, the ability to override any existing pixel, and a 5-minute cooldown between placements. The combination meant that any lasting mark on the canvas would have to be negotiated between participants, leading to the emergence of a multitude of collective creative behaviours. If this simple 5-minute constriction had been reduced to 1 second, the platform would have collapsed into single-user monopolies, and died out. This simple example tells us that we cannot disregard the importance of the medium's affordances in conditioning what creative artefacts are explored and produced \cite{mcluhan1994understanding}.

Existing CC frameworks of creative potential formalise reachability: Boden's \textit{conceptual spaces}~\shortcite{boden92} and Wiggins' \textit{Creative Systems Framework}~\shortcite{wiggins2006preliminary} describe what concepts an agent can reach by searching a given conceptual space. McCormack's \textit{creative ecosystems}~\shortcite{McCormack2012} extend the picture by treating that space as co-evolving with the agent through niche-construction, describing the medium through the joint dynamics of agents acting in it. Our contribution sits one level below both. We propose that a medium's affordances, as a design-level mechanism, induce macroscopic complex-systems properties that belong to the medium itself rather than to any particular agent's strategy within it. Those properties then condition the agent's later search, and the creative artefacts this results in. Because the vocabulary is substrate-agnostic, media with very different internal dynamics (\texttt{r/place}, \texttt{Twitter}, \texttt{Minecraft}) can be studied on the same basis.

\section{Creative Media as Complex Systems}

Complex systems are some of the most interesting phenomena we observe in natural, artificial and social systems: from the swarming of starlings, self-replicating patterns in \texttt{Conway's Game of Life} or the collective canvas of \texttt{r/place}. Behind all of these diverse phenomena, there is a unifying mechanism: many components interacting through simple rules to give rise to emergent behaviours not reducible to any one component in isolation; e.g., individual participants of the \texttt{r/place} experiment cannot know how their pixels will condition the trajectory of the global canvas. These systems are referred to as \emph{complex systems}~\cite{mitchell2009complexity}.

In order to analyse a creative medium as a complex system, we propose to characterise media using a set of complexity properties: emergence, collective intelligence and self-organisation, non-linear dynamics, criticality, multi-scale hierarchy, phase transitions, diversity of attractors, path dependence, and open-endedness. The remainder of this section takes them one at a time, working out for each what structural conditions a medium's affordances must satisfy to produce it and which CC concepts it re-describes in CS vocabulary, with a worked medium example (Figure~\ref{fig:iccc}).

\paragraph{Emergence.} When observed across different scales, certain systems display \textit{macroscopic} patterns that are qualitatively different from the behaviours of each individual component at the \textit{micro-level}. For instance, the macroscopic behaviour of a complex organism appears to transcend the biochemistry that produces it. These emergent phenomena are naturally described using concepts that do not exist in the language of the system's microscopic components: a ``glider'' in \texttt{Game of Life} is a meaningful macro-level concept, yet it has no counterpart in the micro-level language of cell states and neighbourhood rules. While emergent behaviours do follow in principle from the micro-level rules, they are not readily predictable from them, a consequence of non-linear interactions propagating across spatial and temporal scales.

In the CC literature, the micro-macro distinction maps onto Boden's~\shortcite{boden92} separation between \textit{exploratory} and \textit{transformational creativity}, later formalised by Wiggins'~\shortcite{wiggins2006preliminary} separation between emergence \textit{within} a conceptual space $\mathcal{L}$ and emergence \textit{of} a new $\mathcal{L}$. Taking the aforementioned \texttt{r/place} example~\shortcite{rappaz2018latent}: the per-pixel placement and override-anywhere rule are deliberately too coarse to encode any macro object directly, so flags, portraits and alliance territories only become visible at the canvas scale. Reading emergence through complexity science adds a design lever absent from earlier CC framings: the gap between micro and macro vocabularies is something the medium designer can set, by choosing how coarse the affordances are. 

\paragraph{Collective intelligence and self-organisation.} \textit{Collective intelligence} refers to self-organising phenomena where individual agents coordinate and integrate information to solve higher-order problems that are inaccessible to individuals. The underlying mechanism, \textit{self-organisation}, refers to the spontaneous formation of order purely from local interactions: no single element directs the entire system; instead, decision-making and influence are spread across many components, often referred to as \textit{distributed control}~\cite{camazine2020self}. The time evolution toward ordered states, in the form of a fixed point or limit cycle, can be caused by simple negative feedback loops or complex interaction motifs.

CC has engaged with this property under several names: \textit{collective creativity} \cite{sawyer2003group}, \textit{social creativity}~\cite{saunders2015computational,fischer2005beyond}, \textit{creative ecosystems}~\cite{bown2012generative,maher2010evaluating}, and \textit{co-creativity}~\cite{kantosalo2016modes,linkola2019extending}. Using the complexity concept as a recipe, we get a sense of what such a ``coordination layer'' must contain: stateful elements, a notion of locality, and a local-state update mechanism with no central controller. \texttt{Wikipedia's talk pages}~\cite{viegas2007talk} instantiate this set of properties, and produce two creative artefacts in the process: collaboratively written articles, and institutional policies that emerge from practice rather than from any top-down design. The concrete affordances used to realise these properties include threaded replies (locality), signatures plus edit-versioning (per-contributor state), and watchlist-driven attention (decentralised update).

\paragraph{Non-linear dynamics.} Systems exhibit non-linear interactions where small perturbations can produce system-wide effects. These dynamics are typically driven by feedback loops and non-linear interactions that amplify local changes into global effects.

Non-linearity is ubiquitous in CC media but rarely formalised under that name; instead the phenomenon recurs under different headings: \textit{deceptive landscapes} in Stanley and Lehman's novelty search~\shortcite{lehman2011abandoning}, \textit{predictive deviation} in Maher and Grace's surprise search~\shortcite{grace2014expect}, \textit{combinatorial explosion} in Boden's combinatorial creativity~\shortcite{boden92}, and the dynamical-systems work on interactive music ~\cite{eldridge2008collaborating,bown2012generative} that names non-linearity directly. From the complexity concept, what combines all of these is a coupling that ties local actions to global state, and a feedback channel that lets small differences amplify rather than damp out. The microblogging website \texttt{Twitter}~\cite{goel2016structural} instantiates this pattern through the reshare button which couples a local post to global reach (coupling) and the heavy-tailed follower-graph which amplifies small differences in reach into large differences in engagement (feedback).

\paragraph{Criticality and Edge of Chaos.} Some systems sit at a regime between order and disorder, where correlations span all scales and the system is maximally responsive to inputs. At criticality, fluctuations are scale-free and small perturbations can propagate arbitrarily far. The \textit{edge of chaos} hypothesis~\cite{langton1990computation} associates this regime with the richest emergent computation, sitting between frozen, predictable dynamics on one side and disordered, uncorrelated dynamics on the other.

CC literature encounters criticality in Langton~\shortcite{langton1990computation} and Wolfram's class-IV cellular automata~\shortcite{wolfram1984universality}; specially relevant in context of generative-art parameter sweeps~\shortcite{mccormack2001art}, where the ``interesting'' regime is narrow. If we look to complexity science for what characterises this property, we find that we need a tuned \textit{control parameter} that establishes an ordered regime on one side and a disordered regime on the other, and a narrow window between them where small disturbances can propagate across the full system. \texttt{Boids}~\cite{reynolds1987flocks} is an illustrative creative instance: each boid steers by three local rules --- cohesion (steer toward neighbours' centroid), alignment (match neighbours' heading), and separation (avoid crowding) --- whose weights act as the control parameter. Turn cohesion and alignment up and the flock collapses into a tight cluster locked to the average heading (ordered); turn them down and the boids drift independently, one boid's turn never reaching another (disordered). The narrow band between is flocking proper: a single turn ripples across the flock and reshapes it without freezing or shattering it.

\paragraph{Multi-scale hierarchy.} Complex systems often exhibit multiple nested levels of interactions and feedback loops, operating across scales. In many cases, these hierarchies are not imposed, but instead emerge from the lower organisational levels. Interactions among elements within each level, and across different levels, typically occur on distinct time scales.

CC's most explicit treatment of hierarchy is Wiggins' meta-search~\shortcite{wiggins2006preliminary}, where a slow outer loop modifies the rule-set $\mathcal{R}$ which the fast inner loop searches under. Saunders's three-level model~\shortcite{saunders2015computational} (artefact / evaluator / society) and the Csikszentmihalyi-derived individual / domain / field framing~\shortcite{csikszentmihalyi1999implications} also layer creative activity similarly. From the complex system lens, each level has its own vocabulary and its own dynamics, with within-level interactions stronger or faster than across-level coupling (Simon's \textit{near-decomposability}~\cite{simon1962architecture}), so that the level can be reasoned about in its own terms. \texttt{Minecraft}~\cite{duncan2011minecraft} illustrates several such levels at once: blocks, structures, biomes, and world-scale builds nest spatially on the artefact side, players, factions, and server cultures nest socially, and redstone introduces a distinct vocabulary (signals, gates, computation) above the block-placement layer.

\paragraph{Presence of phase transitions.} A system exhibits phase transitions when its macroscopic state changes suddenly and qualitatively as a control parameter (temperature, density, connectivity, etc.) crosses a critical value. Originally formalised in statistical physics, phase transitions are also observed in networks, populations, and information-processing systems, where they typically separate qualitatively distinct macro-regimes.

Phase transitions are not standard CC vocabulary, but the underlying phenomenon is. Boden's \textit{transformational creativity} is structurally a discrete change in $\mathcal{R}$~\shortcite{boden2004creative}, Wiggins' CSF formalises it as a move in the meta-space over $\mathcal{R}$~\shortcite{wiggins2006preliminary}, and Csikszentmihalyi's \textit{creative breakthroughs}~\shortcite{csikszentmihalyi1999implications} occupy the same slot under a systems reading. The complex systems approach decomposes the phenomenon into its ingredients: a tunable control parameter, a critical value of that parameter, and a qualitative regime change in the dynamics across it. Here again the \texttt{r/place} medium makes for a good example. Below a critical canvas density, participants paint freely in uncontested regions, each placement effectively independent of every other. Above it, every placement must overwrite something a prior participant cared about, and the medium switches phase: alliances form, territories are defended, and off-platform Discords organise raids and counter-raids. The bounded canvas makes interaction patterns a function of the cooldown timer (control parameter) and the user's cooldown value results in coordination over monopoly as the post-threshold equilibrium (regime change).

\paragraph{Diversity of attractors.} In dynamical system theory, an attractor is a region of state space toward which trajectories of the system state evolve. The mere presence of attractors says little about a creative medium, since any sufficiently constrained system will exhibit some. What matters for CC is the \textit{diversity} of attractors a medium supports---fixed points, limit cycles, or strange attractors of qualitatively different kinds---whose collective structure bounds the long-term qualitative behaviours the medium can express.

Attractor language is sparse in CC and mostly metaphorically used (\textit{creative ruts}, \textit{style basins}, \textit{mode collapse}), with explicit uses in Eldridge's interactive-music work~\shortcite{eldridge2008collaborating} and Bown on musical agents~\shortcite{bown2012generative}. Two CC concepts touch the diversity aspect: Ritchie's \textit{typicality}~\shortcite{ritchie07}, proximity to a basin of attraction; and Koestler's \textit{bisociation}~\shortcite{koestler1964act}, later formalised by Fauconnier and Turner as \textit{conceptual blending}~\shortcite{fauconnier2002way}, a cross-basin operation that presupposes the substrate hosts multiple kinds of basin to combine across. Fixed points need stable rest configurations the system can settle into; limit cycles need feedback or scheduling structures with a natural period of return; strange attractors need a bounded variation channel rich enough that trajectories stay confined to a regime without ever repeating. The mix a medium supports determines the modes of expression it makes available: a substrate offering only the first sustains only convergence, while one supplying all three sustains convergence, recurrence, and perpetual variation.  An example that demonstrates the notion of attractors is \texttt{Wikipedia's talk pages}~\cite{viegas2007talk}. Godwin's law --- the observation that long-running online debates tend to invoke Nazi comparisons regardless of starting topic --- and articles that have stabilised on an accepted version are the fixed-point case (discussion converges and stays); while debates that flare up every few months on contested topics are the limit-cycle case~\cite{sumi2011edit}.

\paragraph{Path dependence.} The state a system reaches depends not only on its current inputs but on the order in which past events occurred: equivalent boundary conditions reached through different histories lead to different futures. A special case is \textit{stigmergy}, where agents coordinate indirectly by modifying a shared environment, each action leaving a trace that constrains and guides subsequent actions. Path-dependence appears in CC under several names: Stanley and Lehman's \textit{stepping-stone} reading of novelty search~\shortcite{lehman2011abandoning}, where interesting end-points are reached only through paths whose intermediate stages don't look like progress toward them; Tomasello's \textit{cumulative cultural ratchet}~\shortcite{tomasello1999cultural}, in which each generation builds on the artefacts of the last; and lineage-based generative systems such as \texttt{Picbreeder}~\shortcite{secretan2011picbreeder}, where every image is a leaf of a specific evolutionary path. From the CS account, the phenomenon decomposes into two ingredients: a persistent state that records past events, and a coupling between current actions and that state under which each action both reads from it and modifies it. Without persistence there is no path for the future to depend on; without read---modify coupling each action operates independently of history.

An interesting example is the gaming modality, or artistic form, known as \texttt{speedrunning}~\cite{scullyblaker2014practiced}: each newly-discovered glitch becomes a stepping stone for the next, and canonical ``any\%'' routes accumulate from sequences of exploits whose individual stages are not interesting on their own. The community-maintained archive of routes, splits, and frame-counts supplies the state, and the practice of forking an existing route to insert a new exploit supplies the coupling. The stigmergic special case~\cite{bonabeau1999swarm} fits \texttt{r/place}~\shortcite{rappaz2018latent}: the persistent canvas is the state, and override-anywhere plus the absence of direct messaging force coordination to read and modify it.

\paragraph{Open-endedness.} A system is open-ended when it has the capacity to continuously produce novel, increasingly \textit{interesting} and diverse behaviours.

Open-endedness (OE) is the deepest pre-existing bridge between CC and Complexity Science: Stanley and Lehman on open-ended evolution~\shortcite{lehman2011abandoning}; Banzhaf et al.\ on OE criteria; Chan's \texttt{Lenia}~\shortcite{chan2019lenia} and Mordvintsev's \texttt{Particle Lenia} treat OE as a substrate property to be engineered. From the complex systems perspective, OE decomposes into three ingredients: an unbounded reachable set the medium can in principle produce (substrate OE), a novelty-generating dynamic that keeps producing new configurations (process OE, often via coupling between agents and substrate), and the absence of an imposed objective that would close the space by selecting for convergence. \texttt{Picbreeder}~\shortcite{secretan2011picbreeder} instantiates all three: its CPPN encoding makes the image space effectively unbounded (the reachable set); aesthetic selection by users drives the trajectory under their evolving preferences (the `novelty engine'); and the deliberate absence of any fitness function lets lineages diverge into shapes nobody set out to produce (no imposed closure).

\vspace{1em}
Formalising affordances through this conceptual basis of complex systems notions allows creative media to be compared along principled dimensions. Two routes are available. From the affordance side (micro to macro), a medium can be characterised by asking which structural conditions its affordances supply, taking each property's recipe from the previous section as a checklist. From the complex systems perspective (macro to micro), a medium's profile can be assessed directly: either phenomenologically, by observing the medium and judging which properties feel present at what intensity, or formally, drawing on measures from the complexity literature for each properties---e.g., entropy measures, long-range correlation functions for criticality, distance-to-attractor measures for diversity of attractors, network-statistical signatures for collective intelligence, and so on ~\cite{McCormack2022Dec}.

\section{Conclusions}

In this work, we have presented a complex-systems perspective on computational creativity with the goal of establishing a bridge between the complex systems and the computational creativity communities. By formalising computational creativity notions under the complexity science lens, the formalism allows us to explore how the affordances of a medium---thought of as design choices---determine the creative artefacts it produces.

We have focused on nine system-level properties: emergence, collective intelligence and self-organisation, non-linear dynamics, criticality, multi-scale hierarchy, phase transitions, diversity of attractors, path dependence, and open-endedness. 
For each property, we point to the affordances that enable it (i.e., a narrow order-disorder window for criticality, near-decomposable levels with distinct vocabularies for hierarchy, a persistent state with read-modify coupling for path-dependence), and further, we trace these affordances to specific design choices in existing creative media: \texttt{r/place}'s per-user cooldown, the cohesion-alignment-separation weights in \texttt{Boids}, the threading-and-signing of \texttt{Wikipedia talk pages}. The formalism is therefore not only a tool for describing media after the fact, but a causal account of how the choice of affordances (i.e., the micro-level) affects the quality and properties of a medium (i.e., the macro-level). This allows us to work backwards from how we intend the artist to engage with artefact creation, to the design of the medium itself. It also opens a path toward identifying under-explored combinations of complexity-informed properties, and outlining the yet-to-exist creative media they would yield.

\bibliographystyle{iccc}
\bibliography{references}

@book{mcluhan1994understanding,
  title={Understanding media: The extensions of man},
  author={McLuhan, Marshall},
  year={1964},
}

@book{camazine2020self,
  title={Self-Organisation in Biological Systems},
  author={Camazine, Scott and Deneubourg, Jean-Louis and Franks, Nigel R. and Sneyd, James and Bonabeau, Eric and Theraula, Guy},
  year={2001},
}

@article{wiggins2006preliminary,
  author  = {Wiggins, Geraint A.},
  title   = {A preliminary framework for description, analysis and comparison of creative systems},
  journal = {Knowledge-Based Systems},
  volume  = {19},
  number  = {7},
  pages   = {449--458},
  year    = {2006},
  doi     = {10.1016/j.knosys.2006.04.009}
}

@book{boden2004creative,
  author    = {Boden, Margaret A.},
  title     = {The Creative Mind: Myths and Mechanisms},
  edition   = {2nd},
  year      = {2004},
  doi       = {10.4324/9780203508527}
}

@incollection{McCormack2012,
	author = {McCormack, Jon},
	title = {{Creative Ecosystems}},
	booktitle = {{Computers and Creativity}},
	pages = {39--60},
	year = {2012},
	doi = {10.1007/978-3-642-31727-9_2}
}

@inproceedings{rappaz2018latent,
  author    = {Rappaz, J\'er\'emie and Catasta, Michele and West, Robert and Aberer, Karl},
  title     = {Latent Structure in Collaboration: The Case of Reddit r/place},
  booktitle = {Proc. International AAAI Conference on Web and Social Media},
  volume    = {12},
  number    = {1},
  year      = {2018},
  doi       = {10.1609/icwsm.v12i1.15013}
}

@book{mitchell2009complexity,
  author    = {Mitchell, Melanie},
  title     = {Complexity: A Guided Tour},
  year      = {2009},
  url       = {https://global.oup.com/academic/product/complexity-9780199798100}
}

@book{sawyer2003group,
  author    = {Sawyer, R. Keith},
  title     = {Group Creativity: Music, Theater, Collaboration},
  year      = {2003},
  isbn      = {978-0805844351}
}

@book{gibson1979ecological,
  title={The ecological approach to visual perception: classic edition},
  author={Gibson, James J},
  year={1979},
}

@book{boden92,
author = {Margaret Boden},
title = {The Creative Mind},
year = {1992},
}

@article{ritchie07,
 author  = {Graeme Ritchie},
 title = {Some Empirical Criteria for Attributing Creativity to a Computer Program},
 journal = {Minds and Machines},
 year = {2007},
 pages = {76-99},
 volume = {17},
}

@inproceedings{grace2014expect,
  author    = {Grace, Kazjon and Maher, Mary Lou},
  title     = {What to Expect When You're Expecting: The Role of Unexpectedness in Computationally Evaluating Creativity},
  booktitle = {Proc. International Conference on Computational Creativity},
  year      = {2014},
}

@article{lehman2011abandoning,
  author    = {Lehman, Joel and Stanley, Kenneth O.},
  title     = {Abandoning Objectives: Evolution Through the Search for Novelty Alone},
  journal   = {Evolutionary Computation},
  year      = {2011},
  volume    = {19},
  number    = {2},
  pages     = {189--223},
  doi       = {10.1162/EVCO_a_00025},
}

@article{secretan2011picbreeder,
  author    = {Secretan, Jimmy and Stanley, Kenneth O.},
  title     = {Picbreeder: A Case Study in Collaborative Evolutionary Exploration of Design Space},
  journal   = {Evolutionary Computation},
  year      = {2011},
  volume    = {19},
  number    = {3},
  pages     = {373--403},
  doi       = {10.1162/EVCO_a_00030},
}

@article{chan2019lenia,
  author  = {Chan, Bert Wang-Chak},
  title   = {Lenia: Biology of Artificial Life},
  journal = {Complex Systems},
  year    = {2019},
  volume  = {28},
  number  = {3},
  pages   = {251--286}
}

@article{langton1990computation,
  author  = {Langton, Christopher G.},
  title   = {Computation at the edge of chaos: Phase transitions and emergent computation},
  journal = {Physica D: Nonlinear Phenomena},
  year    = {1990},
  volume  = {42},
  number  = {1--3},
  pages   = {12--37},
  doi     = {10.1016/0167-2789(90)90064-V}
}

@article{simon1962architecture,
  author  = {Simon, Herbert A.},
  title   = {The Architecture of Complexity},
  journal = {Proceedings of the American Philosophical Society},
  year    = {1962},
  volume  = {106},
  number  = {6},
  pages   = {467--482},
  url     = {https://www.jstor.org/stable/985254}
}

@book{bonabeau1999swarm,
  author    = {Bonabeau, Eric and Dorigo, Marco and Theraulaz, Guy},
  title     = {Swarm Intelligence: From Natural to Artificial Systems},
  year      = {1999},
  isbn      = {9780195131598}
}

@book{tomasello1999cultural,
  author    = {Tomasello, Michael},
  title     = {The Cultural Origins of Human Cognition},
  year      = {1999},
  isbn      = {9780674005822}
}

@inproceedings{viegas2007talk,
  author    = {Vi{\'e}gas, Fernanda B. and Wattenberg, Martin and Kriss, Jesse and van Ham, Frank},
  title     = {Talk Before You Type: Coordination in {W}ikipedia},
  booktitle = {Proc. Hawaii International Conference on System Sciences},
  year      = {2007},
  doi       = {10.1109/HICSS.2007.511}
}

@article{goel2016structural,
  author  = {Goel, Sharad and Anderson, Ashton and Hofman, Jake and Watts, Duncan J.},
  title   = {The Structural Virality of Online Diffusion},
  journal = {Management Science},
  year    = {2016},
  volume  = {62},
  number  = {1},
  pages   = {180--196},
  doi     = {10.1287/mnsc.2015.2158}
}

@article{duncan2011minecraft,
  author  = {Duncan, Sean C.},
  title   = {Minecraft, Beyond Construction and Survival},
  journal = {Well Played: A Journal on Video Games, Value and Meaning},
  year    = {2011},
  volume  = {1},
  number  = {1},
  pages   = {1--22}
}

@article{scullyblaker2014practiced,
  author  = {Scully-Blaker, Rainforest},
  title   = {A Practiced Practice: Speedrunning Through Space With de {C}erteau and {V}irilio},
  journal = {Game Studies},
  year    = {2014},
  volume  = {14},
  number  = {1},
  url     = {https://gamestudies.org/1401/articles/scullyblaker}
}

@book{koestler1964act,
  author    = {Koestler, Arthur},
  title     = {The Act of Creation},
  year      = {1964}
}

@book{fauconnier2002way,
  author    = {Fauconnier, Gilles and Turner, Mark},
  title     = {The Way We Think: Conceptual Blending and the Mind's Hidden Complexities},
  year      = {2002},
  isbn      = {9780465087860}
}

@inproceedings{sumi2011edit,
  author    = {Sumi, R{\'o}bert and Yasseri, Taha and Rung, Andr{\'a}s and Kornai, Andr{\'a}s and Kert{\'e}sz, J{\'a}nos},
  title     = {Edit Wars in {W}ikipedia},
  booktitle = {Proc. IEEE International Conference on Privacy, Security, Risk and Trust},
  year      = {2011},
  doi       = {10.1109/PASSAT/SOCIALCOM.2011.47}
}

@article{wolfram1984universality,
  author  = {Wolfram, Stephen},
  title   = {Universality and complexity in cellular automata},
  journal = {Physica D},
  year    = {1984},
  volume  = {10},
  number  = {1--2},
  pages   = {1--35},
  doi     = {10.1016/0167-2789(84)90245-8}
}

@incollection{bown2012generative,
  author    = {Bown, Oliver},
  title     = {Generative and Adaptive Creativity: A Unified Approach to Creativity in Nature, Humans and Machines},
  booktitle = {Computers and Creativity},
  year      = {2012},
  doi       = {10.1007/978-3-642-31727-9_14}
}

@article{saunders2015computational,
  author  = {Saunders, Rob and Bown, Oliver},
  title   = {Computational Social Creativity},
  journal = {Artificial Life},
  volume  = {21},
  number  = {3},
  pages   = {366--378},
  year    = {2015},
  doi     = {10.1162/ARTL_a_00177}
}

@article{fischer2005beyond,
  author  = {Fischer, Gerhard and Giaccardi, Elisa and Eden, Hal and Sugimoto, Masanori and Ye, Yunwen},
  title   = {Beyond Binary Choices: Integrating Individual and Social Creativity},
  journal = {International Journal of Human-Computer Studies},
  volume  = {63},
  number  = {4--5},
  pages   = {482--512},
  year    = {2005},
  doi     = {10.1016/j.ijhcs.2005.04.014}
}

@inproceedings{maher2010evaluating,
  author    = {Maher, Mary Lou},
  title     = {Evaluating Creativity in Humans, Computers, and Collectively Intelligent Systems},
  booktitle = {Proc. DESIRE Network Conference on Creativity and Innovation in Design},
  year      = {2010},
  doi       = {10.1145/1854969.1854973}
}

@inproceedings{kantosalo2016modes,
  author    = {Kantosalo, Anna and Toivonen, Hannu},
  title     = {Modes for Creative Human-Computer Collaboration: Alternating and Task-Divided Co-Creativity},
  booktitle = {Proc. International Conference on Computational Creativity},
  pages     = {77--84},
  address   = {Paris},
  year      = {2016}
}

@inproceedings{linkola2019extending,
  author    = {Linkola, Simo and Kantosalo, Anna},
  title     = {Extending the Creative Systems Framework for the Analysis of Creative Agent Societies},
  booktitle = {Proc. ICCC},
  address   = {Charlotte, NC},
  year      = {2019}
}

@misc{eldridge2008collaborating,
  author = {Eldridge, Alice C.},
  title  = {Collaborating with the Behaving Machine: Simple Adaptive Dynamical Systems for Generative and Interactive Music},
  year   = {2008}
}

@incollection{csikszentmihalyi1999implications,
  author    = {Csikszentmihalyi, Mihaly},
  title     = {Implications of a Systems Perspective for the Study of Creativity},
  booktitle = {Handbook of Creativity},
  pages     = {313--335},
  year      = {1999}
}

@inproceedings{mccormack2001art,
  author    = {McCormack, Jon and Dorin, Alan},
  title     = {Art, Emergence and the Computational Sublime},
  booktitle = {Proc. International Conference on Generative Systems in the Electronic Arts},
  address   = {Victoria, Australia},
  year      = {2001}
}

@inproceedings{reynolds1987flocks,
  author    = {Reynolds, Craig W.},
  title     = {Flocks, Herds and Schools: A Distributed Behavioral Model},
  booktitle = {Proc. Conference on Computer Graphics and Interactive Techniques},
  pages     = {25--34},
  year      = {1987},
  address   = {New York, NY, USA},
  doi       = {10.1145/37401.37406}
}

@book{murray2011inventing,
  author    = {Murray, Janet H.},
  title     = {Inventing the Medium: Principles of Interaction Design as a Cultural Practice},
  year      = {2011},
  isbn      = {9780262016148},
  pages     = {483}
}

@article{McCormack2022Dec,
  title={Complexity and aesthetics in generative and evolutionary art},
  author={McCormack, Jon and Cruz Gambardella, Camilo},
  journal={Genetic Programming and Evolvable Machines},
  volume={23},
  number={4},
  pages={535--556},
  year={2022},
  publisher={Springer}
}

\end{document}